\documentclass[prb,aps,twocolumn,floatfix,showpacs]{revtex4}
\usepackage{bm}
\usepackage{amssymb}
\usepackage{graphicx}

\begin{document}

\title{Octupolar ordering of classical kagome antiferromagnets in two and 
three dimensions}

\author{M. E. Zhitomirsky}
\affiliation{Commissariat \`a l'Energie Atomique, DSM/INAC/SPSMS, 
17 r.\ des Martyrs,  F-38054 Grenoble, France}

%%%%%%%%%%%%%%%%%%%%%%%%%%%%%%%%%%%%%%%%%%%%%%%%%%%%%%%%%%%%%%%%%%%%%%%%%
\begin{abstract}
Classical Heisenberg antiferromagnets on two-dimensional kagome and 
three-dimensional hyperkagome lattices are investigated by Monte Carlo 
simulations. For both models the symmetry-breaking states at low 
temperatures are described by non-zero octupole moments or third-rank 
spin tensor order parameters.
In the case of the two-dimensional kagome antiferromagnet, 
a sharp crossover into a coplanar state takes place at 
$T_k\approx 0.004J$, which we attribute to proliferation of fractional 
vortices.
The three-dimensional model exhibits a first-order transition
at $T_c\approx 0.002J$ into a phase with critical spin correlations,
which possesses a long-range order of octupole moments.
\end{abstract}
%%%%%%%%%%%%%%%%%%%%%%%%%%%%%%%%%%%%%%%%%%%%%%%%%%%%%%%%%%%%%%%%%

\date{7 May 2008, revised 8 August 2008}
 
\maketitle

\section{Introduction}

\begin{figure}[t]
\includegraphics[width=0.8\columnwidth]{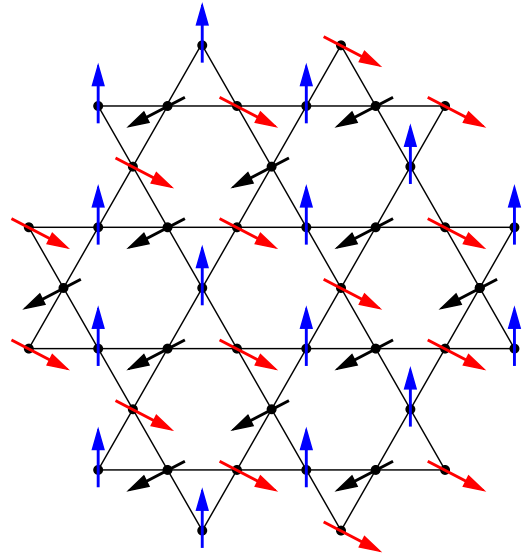}
\caption{(Color online) Section of the kagome lattice with spins in 
the fully ordered $\sqrt{3}\times\!\sqrt{3}$ structure.}
\label{lattice2D}
\end{figure}

A two-dimensional (2D) network of corner-sharing triangles 
known as the kagome lattice, Fig.~\ref{lattice2D}, is a prototype 
of geometrical frustration. The nearest-neighbor Heisenberg  
antiferromagnet on such 
a lattice has an infinite number of spin configurations minimizing
the exchange energy.
Both quantum \cite{misguich04} and classical 
\cite{chalker92,harris92,huse92,ritchey93,chandra93,reimers93,mzh02} 
spin models on the kagome lattice have attracted significant 
theoretical interest in the past. Realizations of the kagome 
lattice topology among magnetic solids were initially rather 
scarce with the prime example being SrCr$_{8-x}$Ga$_{4+x}$O$_{19}$.
\cite{obrados88,ramirez90} 
In the last few years a significant number of new magnetic compounds 
that are believed to be related to the kagome lattice 
antiferromagnet have been synthesized and studied. 
\cite{lee97,wills98,wills00,inami00,hiroi01,grohol03,shores05,%
matan06,schweika07,mendels07,fak08} 
Often these materials suffer from substitutional disorder,
are affected by small structural deviations from the ideal kagome network,
or have extra interactions, which lift the magnetic degeneracy.
Nevertheless, recent neutron scattering experiments on powder samples of
large-$S$ kagome materials, Y$_{0.5}$Ca$_{0.5}$BaCo$_4$O$_7$\, \cite{schweika07}
($S=3/2$)  and  deuteronium jarosite \cite{fak08}  ($S=5/2$),
have demonstrated remarkable similarity between the measured diffuse 
intensities and the Monte Carlo results for the classical 
model. \cite{reimers93}
Motivated by these two seemingly good realizations of 
the classical kagome antiferromagnet, we reinvestigate 
in the present work the finite-temperature properties of this model.
In particular, we consider the angular dependence of
magnetic correlations, which can be measured in neutron-diffraction
experiments on single crystals. 

A second source of motivation is provided by the recent discovery 
of a 3D array of corner-sharing triangles in a spin-1/2 Mott insulator 
Na$_4$Ir$_3$O$_8$. \cite{okamoto07} Due to similarity with its 2D 
counterpart, this lattice structure has been coined a
hyperkagome lattice. A network of triangles with similar 
topology is also know to exist in gadolinium gallium garnet 
Gd$_3$Ga$_5$O$_{12}$\, \cite{kinney79,schiffer94,petrenko98} ($S=7/2$),
whose enigmatic behavior attracted a lot of theoretical
efforts. \cite{petrenko00,mzh03,yavorskii06} 
Though the magnetic properties of both systems may be quite 
distant from those of the nearest-neighbor classical model, 
in the former case due to quantum  effects, 
in the latter material due to strong dipolar interactions, 
it is still important to understand the properties of the classical
antiferromagnet as a starting reference point.
Furthermore, recent Monte Carlo simulations \cite{hopkinson07} have found
evidence for an interesting low-temperature phase transition for 
the hyperkagome antiferromagnet. 

% -------------------------------------------------------------------------
% -------------------------------------------------------------------------

The classical ground states of the Heisenberg kagome lattice antiferromagnet
are derived from the block representation of the spin Hamiltonian
\begin{equation}
\hat{\cal H} = J \sum_{\langle ij\rangle} {\bf S}_i\cdot{\bf S}_j 
= \frac{J}{2}\sum_\triangle 
\bigl({\bf S}_1+{\bf S}_2+{\bf S}_3 \bigr)^2_\triangle  + \textrm{const}\,.
\label{H0}
\end{equation}	
The energy is minimized by any spin configuration, which has
${\bf S}_\triangle = 0$ for every triangular plaquette.
This classical constraint is satisfied for infinitely many 
configurations including planar and nonplanar states. 
Chalker {\it et al.}\cite{chalker92} have argued that coplanar
spin states are selected by thermal fluctuations
via the order by disorder effect. They have also related 
an asymptotic selection of the spin plane with development of 
the nematic order \cite{andreev84,chandra91} in spin chiralities 
defined as 
\begin{equation}
\mbox{\boldmath $\kappa$} = \frac{2}{3\sqrt{3}} \Bigl( {\bf S}_1\times{\bf S}_2+
{\bf S}_2\times{\bf S}_3+{\bf S}_3\times{\bf S}_1 \Bigr) \ 
\label{chirality}
\end{equation}
for each triangular plaquette.
The published Monte Carlo data seem to confirm this prediction
\cite{chalker92,reimers93} and the corresponding point of view 
prevails now in the literature on frustrated magnets.

Below we present arguments that such a description is {\it incomplete}
and the low-temperature state of the classical 
kagome antiferromagnet should be described by 
a third-rank tensor or octupolar order parameter.
Such a proposal was first put forward a long time ago, 
\cite{ritchey93,chandra93,mzh02}
though no numerical results were presented to substantiate this idea.
The difference between the broken symmetries for the two types of order
parameters is important for topological classification
of point defects in the kagome antiferromagnet.
Topologically stable defects or vortices play a significant role 
in low temperature transformations of 2D geometrically frustrated 
magnets  and may lead to topological phase transitions \cite{kawamura84} 
and/or to an unconventional spin-glass behavior.\cite{ritchey93,chandra93}
The analytic consideration is supported in the following  by
extensive Monte Carlo simulations.

The paper is organized as follows. 
In Sec.~II the possible tensor order parameters are considered
for magnetically disordered spin systems 
and presence of fractional vortices is emphasized in the case of
the classical kagome antiferromagnet.
Section III is devoted to
Monte Carlo results for the 2D kagome antiferromagnet.
In particular, the specific heat exhibits a sharp kink,
which signifies formation of the coplanar spin state.
In Sec.~IV we investigate the behavior of the 3D kagome antiferromagnet
and find that it shows a fluctuation-driven first-order transition. 
The low-temperature phase possesses
no long-range antiferromagnetic correlations and is described instead 
by an  octupolar order parameter.

\section{Spin Tensor Order Parameters}

\subsection{Symmetry analysis}

Let us first recall the arguments for the order 
by disorder effect in the kagome antiferromagnet.
\cite{chalker92,huse92,ritchey93}
Similar discussion for the hyperkagome antiferromagnet
is postponed until Sec. IV.
The classical constraint ${\bf S}_\triangle = 0$ for one triangular 
plaquette is satisfied by a $120^\circ$ spin structure. Once the orientation 
of the three sublattices is chosen for the first plaquette, all lowest 
energy coplanar configurations can be identified with the ground states 
of the three-state Potts antiferromagnet or, equivalently, with coloring 
all sites of the kagome lattice into three colors, such that no two 
neighboring sites have the same color. There are $1.13471^N$ such 
states with $N$ being the number of lattice sites. \cite{baxter70}
The manifold of coplanar states contains two simple periodic structures: 
the $q=0$ state, in which all triangles pointing up ($\triangle$) or 
down ($\bigtriangledown$) 
are in the same state, and the $\sqrt{3}\times\!\sqrt{3}$ structure, 
which is shown in Fig.~\ref{lattice2D} and is described by the wave-vector 
${\bf Q} = (4\pi /3a,0)$, $a$ being the lattice constant. The $q=0$ 
state has the same chiralities  for up and down triangles, while in 
the $\sqrt{3}\times\!\sqrt{3}$ structure chiralities alternate between 
$\triangle$ and $\bigtriangledown$ plaquettes.

Nonplanar states are constructed from planar configurations by 
identifying various closed or open two-color lines. In the 
$\sqrt{3}\times\!\sqrt{3}$ state, these are represented by hexagonal loops,
see Fig.~\ref{lattice2D}.
Nearest neighbors off such a line are necessarily spins of the third color.
Spins on the line can be continuously rotated about the direction
determined by the third sublattice. The obtained spin fold (also called weather-vane mode)
retains the $120^\circ$ spin orientation and costs, therefore, 
no energy. Rotation by  $\pi$ returns spins back into a single plane
creating a  new coplanar state.

The harmonic analysis \cite{chalker92,ritchey93} indicates that 
coplanar states are selected at low temperatures because they have 
the largest number of soft excitations. 
The harmonic excitation spectra are identical for all coplanar
configurations. Hence, selection of a specific translational
pattern, if any, occurs due to weaker nonlinear effects.
Therefore, there should be a range of temperatures where
selection of the spin plane is not accompanied by a wave-vector
selection. The heat capacity in this regime is equal to 
$C = 11/12$ per spin. \cite{chalker92}  
The spin plane is specified by its normal (\ref{chirality}),
which for a general chirality-disordered coplanar state 
selects a line without direction. The corresponding order parameter
is a second-rank traceless tensor: \cite{andreev84,chandra91}
\begin{equation}
Q_\kappa^{\alpha\beta} = \frac{1}{N_\triangle} \sum_p \Bigl(
\kappa_p^\alpha \kappa_p^\beta - \frac{1}{3}\, 
\mbox{\boldmath $\kappa$}_p^2\,\delta_{\alpha\beta}\Bigr)\ ,
\label{quadrupole1}
\end{equation}
where summation extends over all triangular plaquettes.
A simpler form of the nematic order parameter
can be constructed as a sum of on-site quadrupole moments:
\begin{equation}
Q^{\alpha\beta} = \frac{1}{N} \sum_i \Bigl( S^\alpha_i S^\beta_i 
- \frac{1}{3}\, \delta_{\alpha\beta} \Bigr) \ .
\label{quadrupole}
\end{equation}
The two order parameters [Eqs.~(\ref{quadrupole1}) and (\ref{quadrupole})] 
describe the same type of broken symmetry and it is only 
a matter of convenience to choose one of them.

This is not, however, the end of the story. The coplanar 
states break, in addition, the spin-rotational
symmetry inside the plane: At large distances spins do not follow
any specific translational pattern but still are chosen from
the initial sublattice triad. The ground states of the $XY$ kagome antiferromagnet 
with planar spins ${\bf S}_j = (\cos\theta_j,\sin\theta_j)$
have  a long-range order in $w_j = \exp(3i\theta_j)$. \cite{huse92}
Generalization to Heisenberg spins is given by 
an on-site octupole moment expressed as a symmetric third-rank tensor:
\begin{equation}
T_i^{\alpha\beta\gamma} = S^\alpha_i S^\beta_i S^\gamma_i
- \frac{1}{5}\, S_i^\alpha \delta_{\beta\gamma}
- \frac{1}{5}\, S_i^\beta  \delta_{\alpha\gamma} 
- \frac{1}{5}\, S_i^\gamma \delta_{\alpha\beta} \ 
\label{octupoleM}
\end{equation}
with vanishing trace over any pair of indexes. The 
uniform long-range order of such 
octupoles is described by non-zero values of 
\begin{equation}
T^{\alpha\beta\gamma} =  \frac{1}{N} \sum_i \langle 
T_i^{\alpha\beta\gamma} \rangle \ ,
\label{octupole}
\end{equation}
where $\langle\cdots\rangle$ denotes thermodynamic averaging.
The tensor  $T^{\alpha\beta\gamma}$ has in total seven
independent components, as follows from its symmetry and 
tracelessness. 
Note, that a similar duality in the choice between the two tensor order
parameters exists for liquid crystals consisting
of bent-core molecules. \cite{radzihovsky01,lubensky02} 
A complete characterization of the orientational
order in such systems requires definition of a third-rank 
tensor order parameter in addition to the more familiar nematic tensor.
Different forms of third-rank spin tensors
have been discussed in the literature. 
\cite{ritchey93,marchenko88,mzh02,momoi06}
For classical spins all of them are equivalent to
Eq.~(\ref{octupoleM}), the latter form being more convenient 
for numerical simulations.

The order parameters $T^{\alpha\beta\gamma}$ and $Q^{\alpha\beta}$
transform according to different irreducible representations of 
the rotation group SO(3) corresponding to the angular momenta $l=3$ and $l=2$, 
respectively. This does not mean, however, that octupole and
quadrupole moments cannot coexist below the critical point.
The two order parameters
are coupled by a rotationally invariant term
in the free-energy functional
\begin{equation}
\Delta{\cal F}_{QT} \simeq  Q^{\alpha\beta}T^{\alpha\mu\nu}T^{\beta\mu\nu} \ .
\label{FQT}
\end{equation}
Due to time-reversal symmetry, the octupole moment
has zero average value in the nematic phase.
In contrast, an instability driven by $T^{\alpha\beta\gamma}$ 
also induces a nonvanishing quadrupolar tensor unless
$\langle T^{\alpha\mu\nu}T^{\beta\mu\nu}\rangle = \delta_{\alpha\beta}$.
Since the ensemble of coplanar state is described by nonzero
values of both tensors, 
the primary order parameter for
the kagome antiferromagnet is the octupole moment (\ref{octupole}). 
The quadrupole moment in this case is only a secondary order parameter.
In the nomenclature of phase transition theory, 
the low-$T$ state of the kagome antiferromagnet
can be called `{\it improper spin nematic}.'
Numerical data in support of the above conclusion are presented 
in Sec.~IIIB.

Possible symmetries of a uniform octupolar state are deduced
by minimizing the Landau free-energy functional:
\begin{eqnarray}
{\cal F}_T & = & r\, T^{\alpha\beta\gamma} T^{\alpha\beta\gamma}
+ u\, \bigl(T^{\alpha\beta\gamma} T^{\alpha\beta\gamma}\bigr)^2
\nonumber \\
& & \mbox{} 
+ v\, T^{\alpha\beta\gamma} T^{\alpha\delta\mu} T^{\beta\delta\nu} T^{\gamma\mu\nu} \ .
\label{FT}
\end{eqnarray}
For $v>0$, the stable phase has $D_3$ symmetry, whereas for $v<0$, it is invariant 
under tetrahedral point group.\cite{radzihovsky01,lubensky02}
The low-temperature state of the kagome antiferromagnet is, naturally, identified
with  the $D_3$-symmetric ``triatic'' state:
$D_3$ spin rotations permute three spin sublattices and
transform one translationally
disordered coplanar state into another one
from the same ensemble.
The spin tensor for the triatic state
can be parametrized as
\begin{equation}
 T^{\alpha\beta\gamma} \propto \bigl(l^\alpha l^\beta l^\gamma -
 l^\alpha m^\beta m^\gamma - m^\alpha l^\beta m^\gamma - 
 m^\alpha m^\beta l^\gamma \bigr) 
\end{equation}
with two orthogonal unit vectors $\bf l$ and $\bf m$ lying in the spin
plane.
The ``tetrahedratic'' phase ($v<0$) is not realized in the present
spin model. Note, that the interaction term (\ref{FQT}) vanishes in the
tetrahedratic state. Hence stabilization of the triatic phase
can be ascribed to a strong coupling between the octupolar and
the nematic order parameters in the kagome antiferromagnet.

For 2D Heisenberg antiferromagnets the true long-range 
order is impossible at any finite temperature. 
The above discussion applies in this situation
to a symmetry of spin correlations at short distances.
For distances larger than the correlation length $r>\xi$
the order parameters (\ref{quadrupole1})--(\ref{octupole})
vanish and one has to consider instead the generalized susceptibilities
$\chi_V = N\langle V^2\rangle/T$ with $V = Q^{\alpha\beta}$ and 
$T^{\alpha\beta\gamma}$.
The lattice-averaged squares of the two spin tensors,
which are directly measured
in Monte Carlo simulations, are
\begin{equation}
\bigl(Q^{\alpha\beta}\bigr)^2 = \frac{1}{N^2} \sum_{i,j}
\Bigl[\langle ({\bf S}_i\cdot{\bf S}_j)^2\rangle -\frac{1}{3}\Bigr]
\end{equation}
and 
\begin{equation}
\bigl(T^{\alpha\beta\gamma}\bigr)^2 = \frac{1}{N^2} \sum_{i,j} 
\Bigl[ \langle ({\bf S}_i\cdot{\bf S}_j)^3\rangle  - 
\frac{3}{5}\langle {\bf S}_i\cdot{\bf S}_j\rangle  \Bigr].
\end{equation}
At zero temperature, 
in the fully ordered
triatic phase (the ground state of the three-state Potts model),
the above expressions
yield the following limiting values
$\langle (Q^{\alpha\beta})^2 \rangle = 1/6$ and 
$\langle (T^{\alpha\beta\gamma})^2 \rangle = 1/4$ .

\subsection{Topological analysis}

\begin{figure}[t]
\includegraphics[ width = 0.99\columnwidth]{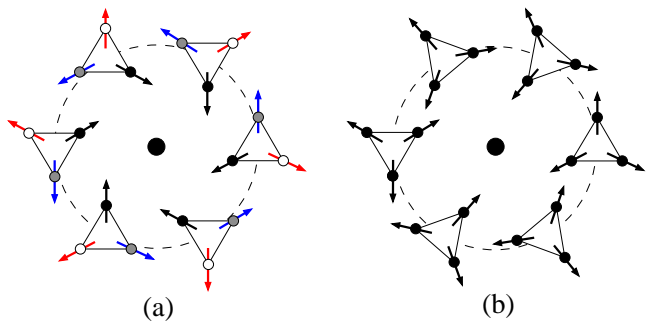}
\caption{(Color online) Stable topological defects in 
noncollinear Heisenberg antiferromagnets.
The three-sublattice $120^\circ$ spin structure is locally
represented by orientations of
equilateral triangles.
Two examples of the spin texture correspond to
(a) $Z_2$-vortex in 
the triangular lattice antiferromagnet with the winding number $n=+1$ and 
(b) fractional vortex in the kagome antiferromagnet with
$n=+1/3$.
}
\label{vortex}
\end{figure}

A state with ordered octupole moments also exhibits nontrivial 
topological properties. \cite{mermin79} 
The order-parameter space of the kagome antiferromagnet
is obtained as a coset space $R=\mathrm{SO(3)}/D_3$.
It has the non-Abelian homotopy  
group $\pi_1(R) = D_6$. The symmetry of 
the octupolar phase allows, therefore, the presence of 
stable topological defects.
It is instructive to compare the topological properties of 
the kagome antiferromagnet with those of its weakly frustrated 
counterpart, the triangular lattice antiferromagnet.
The latter spin system is conventionally ordered into 
the $120^\circ$ spin state, which
breaks completely the rotational symmetry.
The long-range translational order leaves no
discrete symmetries of the spin structure.
As a result, the degeneracy space of the order parameter is $R = \mathrm{SO(3)}$ 
and point defects are $Z_2$-vortices:
$\pi_1[\mathrm{SO(3)}]=\mathbb{Z}_2$. \cite{kawamura84}
A possible realization of $Z_2$-vortex is shown in Fig.~\ref{vortex}a
and corresponds to a pattern  with
all spins lying in a common plane and performing a $2\pi$
rotation around the vortex center.
Equivalently, it can be represented as rotation of equilateral 
triangles with distinguishable vertices.

% -------------------------------------------------------------------------
% -------------------------------------------------------------------------

For a coplanar state of 
the kagome antiferromagnet,
spin folds permute three spin sublattices
and the slow spatial variations of the magnetic order parameter
are represented by a texture of equilateral 
triangles with equivalent vertices.
An elementary point defect is, consequently, 
a fractional vortex 
with $\pm 2\pi/3$ rotation around the vortex core, see Fig.~\ref{vortex}b.
These $1/3$-vortices are known to reduce substantially the Kosterlitz-Thouless 
transition temperature in the $XY$ kagome antiferromagnet.
\cite{rzchowski97,korshunov02}

In Heisenberg magnets topological defects may have
a more complicated nonplanar structure.
Simple hydrodynamic arguments suggest that the
nonplanar $Z_2$-vortex 
in the triangular antiferromagnet has a lower 
energy compared to the planar vortex in Fig.~\ref{vortex}a.
\cite{kawamura84} The same line of arguments applies also 
to defects in the kagome antiferromagnet.
In addition, the hydrodynamic energy of 1/3-vortices 
is further reduced by a factor of $1/9$ due to the smaller phase winding.
As a result, entropic generation of fractional vortices in the kagome 
antiferromagnet starts at significantly lower temperatures than 
a similar effect  for $Z_2$-vortices
in the triangular antiferromagnet.
A possible role of the non-Abelian topological defects
in the kagome antiferromagnet was brought
to attention in Ref.~\onlinecite{ritchey93}
and will be further discussed in Sec.~IIIB.

\section{Kagome Antiferromagnet}

\subsection{Monte Carlo algorithm}

The published Monte Carlo data for the nearest-neighbor 
Heisenberg antiferromagnet on the kagome lattice
were performed on relatively small
clusters of $N=3L^2$ spins with $L\leq 24$ ($N\leq 1728$).
\cite{chalker92,huse92,reimers93} 
Numerical results in the present work have been obtained
for a substantially wider range of lattices
with $L=12$--$72$. 
The standard Metropolis algorithm has been adopted.  
A site 
on a periodic cluster
is randomly picked up and a new orientation of spin is 
chosen. The new direction is accepted according to the Metropolis 
rejection scheme. To increase acceptance rate 
a maximum variation $\Delta S^z=T$ on a $z$-component of spin in 
the local coordinate frame is imposed at low temperatures. 
In this way the acceptance rate stays close to 
50\% in the whole temperature range. 
A sweep over 
the lattice in which on average every spin is attempted to move  
corresponds to one Monte Carlo (MC) step. 

For the Heisenberg kagome antiferromagnet the slowing down 
develops into a serious problem at low temperatures $T/J < 0.01$. 
The autocorrelation time can be further reduced by using 
the microcanonical over-relaxation algorithm. \cite{creutz87} 
Generally, for Heisenberg models the over-relaxation move consists 
in successive rotations of spins around their respective local field 
by an arbitrary angle such that the total energy remains unchanged. 
The simplest and most efficient realization corresponds to 
a $\pi$ rotation, {\it i.e.}, flipping a
spin to the most distant direction from the initial one.\cite{loison05}
Implementation of such a spin move requires neither generation of 
random numbers nor calculation of trigonometric functions, which saves 
significantly operation time. Lattice scans can be performed with 
random or sequential selection of spins. We find that
collective motion of spins is generated more efficiently 
in the latter procedure yielding faster decorrelation.
Finally, one hybrid MC step consists of one canonical MC step
followed by a few microcanonical nonrandom updates. 
Such deterministic reshuffling of spins 
is essential for reducing autocorrelation times at low temperatures.
Typically we use between 3--10 over-relaxation updates per one MC step
depending on cluster size. 

% -------------------------------------------------------------------------
\begin{figure}[b]
\includegraphics[width=0.85\columnwidth]{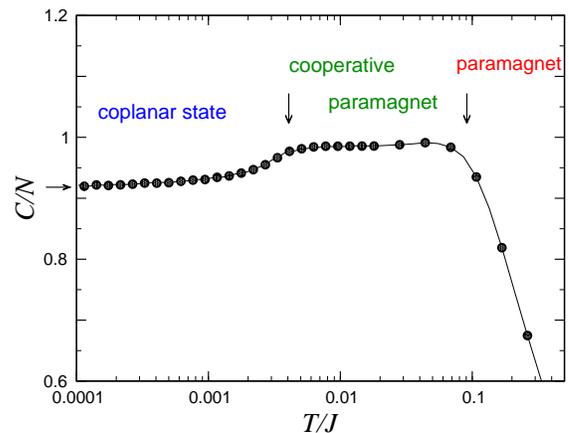}
\caption{(Color online) Temperature dependence of the specific heat 
for a kagome lattice cluster with $L=36$. % ($N=3888$). 
The horizontal arrow denotes the value $C/N=\frac{11}{12}$, the two vertical arrows
indicate boundaries between three different regimes.
}\label{specific1}
\end{figure}
% -------------------------------------------------------------------------

Each finite cluster was initiated with a random spin configuration
and gradually cooled to the lowest temperature $T/J=10^{-4}$.
At every temperature $5\times 10^4$ hybrid MC steps were allowed
for equilibration which were followed by measurements ($\sim 5\times 10^5$)  
performed in intervals of five hybrid MC steps.
In addition, all measured quantities have been averaged 
over 20--50 cooling runs, starting from different random
configurations. This further helps to overcome a freezing problem
and also provides an unbiased
estimate for the statistical errors. Unless otherwise specified,
the error bars do not exceed the symbol sizes.
Special checks have been performed to verify that the hybrid MC algorithm
works efficiently in the relevant temperature range
when instead of gradual cooling we start from either
a random spin configuration or the ordered 
$\sqrt{3}\times\!\sqrt{3}$ structure. Full thermal equilibration of
the $q=0$ ground state was achieved only for $T/J\agt 0.002$, 
which is still significantly better than in the previous studies.
\cite{chalker92,reimers93}

\subsection{Macroscopic properties}

Let us begin with 
the heat capacity, which has been computed  
from fluctuations of the internal energy 
$C = (\langle E^2\rangle - \langle E\rangle^2)/T^2$. 
The temperature dependence of $C(T)$ on a linear-logarithmic
scale is shown in Fig.~\ref{specific1} for a cluster with $L=36$.
One can clearly distinguish three different regimes for the specific heat 
with the two crossover points indicated by vertical arrows.
The high-temperature regime $T/J\agt 0.1$ corresponds to a paramagnetic
phase with only weak correlations between neighboring spins.
In the intermediate regime $0.005 \alt T/J \alt 0.1$ 
the internal energy reaches its classical minimum value $E/N=-J$ up 
to a small contribution from thermal fluctuations. Spins 
on triangular plaquettes become strongly correlated and satisfy 
approximately the constraint condition ${\bf S}_\triangle = 0$.
This regime is commonly known as a classical spin-liquid
or a cooperative paramagnet. \cite{villain79}
The specific heat in the cooperative paramagnetic state 
remains close to $C/N = 1$, which reflects absence of soft modes 
in the excitation spectrum. 

% -------------------------------------------------------------------------
\begin{figure}[t]
\includegraphics[ width = 0.9\columnwidth]{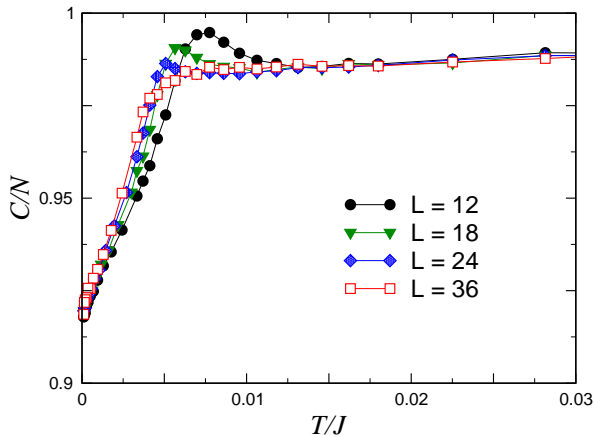}
\caption{(Color online) Finite-size behavior of the specific heat in 
the low temperature region. 
 } \label{specific2}
\end{figure}
% -------------------------------------------------------------------------

Selection of smooth, locally coplanar
spin configurations takes place at $T/J\alt 0.005$ as  
indicated by a reduced specific heat. 
The probability distribution peaks
in the vicinity of coplanar ground states,
which have one zero (anharmonic) mode for every hexagon.
The limiting value
$C/N|_{T\rightarrow 0}$ coincides quite accurately with
$11/12 = 0.916 \ldots$ predicted 
by the mode counting analysis.\cite{chalker92}

The enlarged low-temperature part of $C(T)$ is shown in Fig.~\ref{specific2} 
for several cluster sizes. Two features are noteworthy. First, 
the crossover between a planar spin state and a cooperative paramagnet 
corresponds to a rather sharp kink in $C(T)$ at around $T_k/J = 0.004\pm 0.0005$. 
At $T<T_k$ the specific heat grows linearly with temperature,
which can be accounted for by interaction between the spin waves. 
Second,  the specific heat exhibits a peculiar  finite-size behavior 
in the vicinity of the kink point, showing a rounded
peak on small clusters, which disappears for $L\geq 30$
with no significant finite-size corrections afterward.  

% -------------------------------------------------------------------------
\begin{figure}[b]
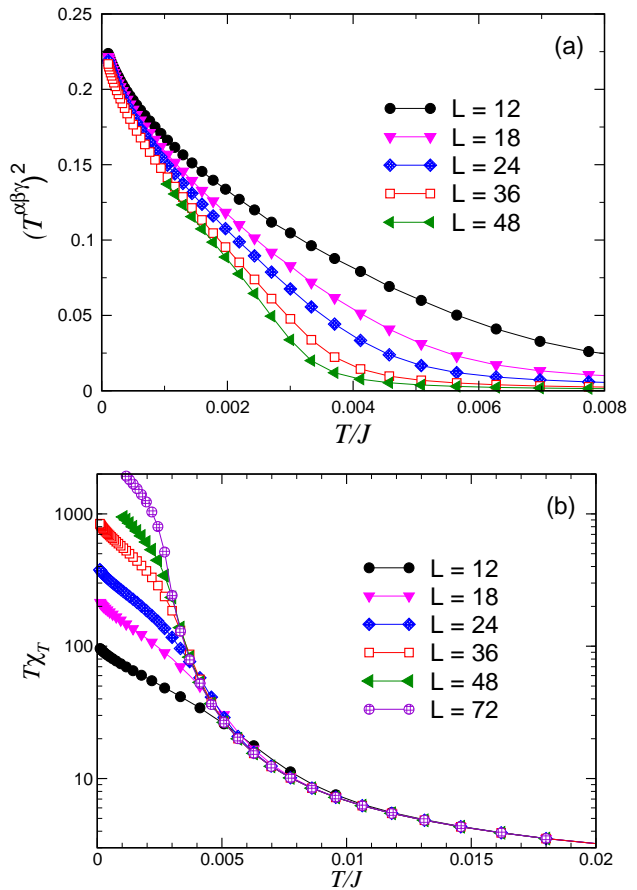

\includegraphics[ width = 0.95\columnwidth]{triatic.eps}
\vspace{3mm}

\includegraphics[ width = 0.91\columnwidth]{chi_triat.eps}
\vspace{3mm}

\caption{(Color online) Temperature dependences of (a) the octupolar
order parameter and (b) the corresponding susceptibility 
for different cluster sizes. 
}
\label{octupolar}
\end{figure}
% -------------------------------------------------------------------------

The temperature dependence of the mean-square of the octupole moment is shown in 
Fig.~\ref{octupolar}a. Large clusters exhibit a clear enhancement of 
the order parameter below $T_k/J \approx 0.004$, which coincides with a kink
position in the specific heat. At low temperatures 
$\langle (T^{\alpha\beta\gamma})^2\rangle$ approaches $1/4$, 
which is the limiting value for the  fully ordered coplanar phase.
The octupolar susceptibility 
\begin{equation}
\chi_T =  \frac{1}{TN} \sum _{i,j} \:
\langle T_i^{\alpha\beta\gamma} T_j^{\alpha\beta\gamma} \rangle 
\label{chiT}
\end{equation}
is presented in Fig.~\ref{octupolar}b.
For each cluster there is an inflection point $T^*_L$ below which 
the correlation length $\xi_T$ becomes of the order of the linear lattice 
size $\xi_T\sim L$ and the susceptibility begins to exhibit 
finite-size effects. The lattice independent part of $\chi_T$ diverges
as $T\rightarrow 0$ signaling a long-range ordered state at $T=0$.
The fast increase in $\chi_T(T)$ at low temperatures is consistent 
with a typical divergence $\chi_T(T) \simeq AT^n\exp(B/T)$ 
found from the nonlinear sigma model mapping, \cite{azaria92}
though no specific predictions for $A$, $B$, and $n$ exists for the 
kagome antiferromagnet.
Note, that a rapid crossover in the behavior of $\chi_T(T)$ 
takes place  in the vicinity of $T/J\sim 0.005$.

\begin{figure}[t]
\includegraphics[ width = 0.85\columnwidth]{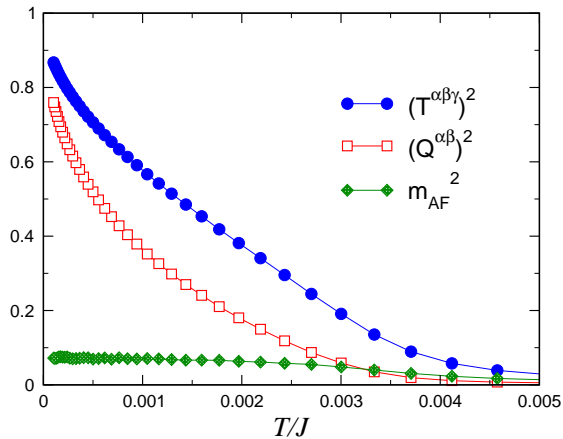}
\caption{(Color online) Temperature dependence 
of the three order parameters normalized
to their respective values in the fully ordered state. Numerical data are 
for a cluster with $L=36$.
}\label{comparisonTQ}
\end{figure}

The behaviors of the two order
parameters $T^{\alpha\beta\gamma}$ and $Q^{\alpha\beta}$ 
is compared in Fig.~\ref{comparisonTQ}.
The octupole moment shows a faster growth with 
decreasing temperature, which would correspond to
a larger exponent if a second-order transition 
is assigned to $T_k$. Fig.~\ref{comparisonTQ}
illustrates our previous conclusion that the octupolar order
parameter drives the low-temperature transformation
in the kagome antiferromagnet.

% -------------------------------------------------------------------------
% -------------------------------------------------------------------------

Finally, we investigate the elastic properties of the kagome antiferromagnet 
by computing the temperature dependence of the spin stiffness.
The spin stiffness $\rho_s$ is defined as the second derivative of
the free-energy with respect to weak nonuniform twist of spins
performed about a certain direction $\alpha$ in spin space:
\begin{equation}
\Delta F = \frac{1}{2} \int d^2r\, \rho_s 
[\mbox{\boldmath $\nabla$}\theta^\alpha({\bf r})]^2 \ .
\end{equation}
Substituting $\theta_i^\alpha=\delta\theta(\hat{\bf e}\cdot{\bf r}_i)$ 
for the twist angle  and taking the limit 
$\delta\theta\rightarrow 0$ we obtain the following expression after
proper symmetrization and normalization per unit area:
\begin{equation}
\rho_s = -\frac{\sqrt{3}}{2N}\Bigl\{ \frac{1}{3}\langle E\rangle+
\frac{J^2}{T} \Bigl\langle \Bigl[ \sum_{\langle ij\rangle}
({\bf S}_i\times{\bf S}_j)^\alpha
(\hat{\bf e}\cdot\mbox{\boldmath $\delta$}_{ij}) \Bigr]^2 \Bigr\rangle
\Bigr\} ,
\end{equation}
where $\langle E\rangle$ is the internal energy and
$\hat{\bf e}$ is an arbitrary unit vector in the lattice plane.

% -------------------------------------------------------------------------
\begin{figure}[b]
\includegraphics[ width = 0.9\columnwidth]{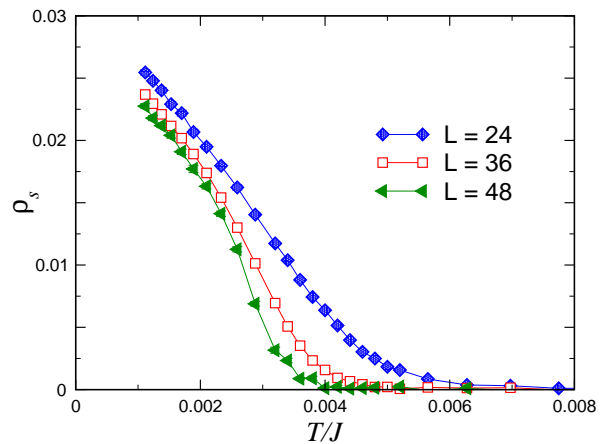}
\caption{(Color online) Spin stiffness of the kagome antiferromagnet
for different cluster sizes.
 } \label{stiff}
\end{figure}
% -------------------------------------------------------------------------

Numerical results for $\rho_s(T)$ are presented in Fig.~\ref{stiff}
for three system sizes. The spin stiffness vanishes at temperatures above
$T/J\sim 0.005$. This further supports identification
of the intermediate phase at $0.005\alt T/J\alt 0.1$ with the cooperative
paramagnet, which has well developed local spin correlations
but exhibits zero response to long-wave-length perturbations.
The spin rigidity starts to increase at $T/J \alt 0.004$ simultaneously
with the development of short-range octupolar correlations. 
Finite-size scaling in the low temperature regime still yields
$\rho_s=0$ as it should be for a 2D Heisenberg spin system.

The origin of the sharp crossover 
in various properties of the 
kagome antiferromagnet at $T\sim T_k \approx 0.004J$ 
deserves special attention. 
The possible phase transition in 2D continuous non-Abelian models
driven by nontrivial topological defects has been discussed
in the context of two different physical applications. 
The first group of works motivated by investigation 
of liquid crystals has studied
the $RP^2$ model in 2D, which in the spin language corresponds
to a model of three-component spins on a square lattice 
coupled with ferro-biquadratic exchange. 
\cite{solomon81,solomon82,kunz92,sanchez03}
The order parameter space is the projective sphere $RP^2 = S^2/Z_2$ 
with the first homotopy group $\pi_1(RP^2) = \mathbb{Z}_2$. 
The topological defects in this context are called disclinations.
Independently, the role of topological defects was emphasized
for 2D noncollinear Heisenberg antiferromagnets.
\cite{kawamura84,wintel94,southern95,caffarel01,kawamura07,domenge} 
The order parameter space is $\mathrm{SO(3)}=S^3/Z_2$ in this case and
the fundamental group is the same $\pi_1[\mathrm{SO(3)}] = \mathbb{Z}_2$. 
For both types of models a straightforward generalization of the 
Kosterlitz-Thouless  scenario suggests that 
topologically stable $Z_2$-vortices are
bound in pairs for $T<T_v$ and become 
free in the high-temperature phase.
\cite{solomon81,kawamura84}  

Kawamura and Miyashita  \cite{kawamura84}
investigated numerically the Heisenberg antiferromagnet on 
a triangular lattice and found evidence for the vortex unbinding 
transition at $T_v\sim 0.3J$. 
The heat capacity exhibits a weak maximum 
in the vicinity of $T_v$. 
The main difference with the standard Kosterlitz-Thouless  transition in  
planar spin systems is that the correlation length remains finite both 
above and below $T_v$.
This leads to a small finite density of free defects
in the low temperature phase.
The low- and the high-temperature phases are still distinguished by
an asymptotic behavior of the vorticity on a large closed contour:
the vorticity function changes from the perimeter law at $T<T_v$ to 
the area law at $T>T_v$. \cite{kawamura84,southern95,kawamura07}
The renormalization group analysis becomes, however, notoriously difficult
since in this case it must include simultaneously
spin-waves and $Z_2$-vortices.
The precise form of a singularity in the thermodynamic potential
at such a topological transition remains unknown up to now. \cite{caffarel01}

The topological properties of the kagome antiferromagnet 
suggest a natural interpretation of the observed crossover
in terms of unbinding of fractional vortices. It 
may also provide another example of topological
transition in 2D Heisenberg antiferromagnets.
The kink anomaly in $C(T)$ is consistent with a cusp-type singularity in
the specific heat found at the topological transition for the $RP^2$
model \cite{kunz92} and for the triangular antiferromagnet.\cite{kawamura84}
The behavior of the spin stiffness also agrees with 
the defect unbinding scenario. Similar to the Kosterlitz-Thouless
transition, free topological defects are responsible
for vanishing $\rho_s=0$  above the crossover point, whereas a much
slower decrease in $\rho_s$ with the system size at low temperatures
is determined by spin-wave excitations.
Further numerical studies, which directly measure 
the density of fractional vortices and the corresponding vorticity function,
are necessary to clarify the above conjecture of topological
transition in the kagome antiferromagnet.

\subsection{Spin correlations}

% -------------------------------------------------------------------------
\begin{figure}[t]
\includegraphics[ width = 0.95\columnwidth]{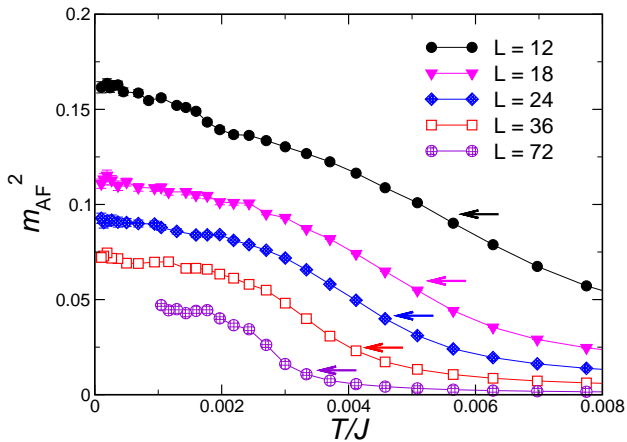}
\caption{(Color online) Temperature dependence of the antiferromagnetic
order parameter for different cluster sizes. Horizontal arrows
indicate corresponding values in the ground state of 
the three-state Potts model.
}
\label{maf}
\end{figure}
% -------------------------------------------------------------------------

The high-temperature series expansion for the kagome 
antiferromagnet\cite{harris92} finds that the maximum in the 
momentum-dependent susceptibility corresponds to the 
$\sqrt{3}\times\!\sqrt{3}$\ spin structure. 
A similar conclusion has been made
by Huse and Rutenberg \cite{huse92} from a  different
perspective: spin correlations of the three-state Potts model
are dominated by the staggered component
at the wave-vector of the $\sqrt{3}\times\!\sqrt{3}$\ structure
with a power-law decay $\sim r^{-4/3}$ at long distances.
Such a purely entropic effect derived from
the mapping to the two-component height model
is related to the fact that the `flat'
$\sqrt{3}\times\!\sqrt{3}$ structure maximizes the number 
of flippable loops. An enhancement of the 
antiferromagnetic correlations at low temperatures
was also seen in the Monte Carlo simulations of the Heisenberg model. 
\cite{huse92,reimers93}

% -------------------------------------------------------------------------
\begin{figure}[b]
\includegraphics[ width = 0.85\columnwidth]{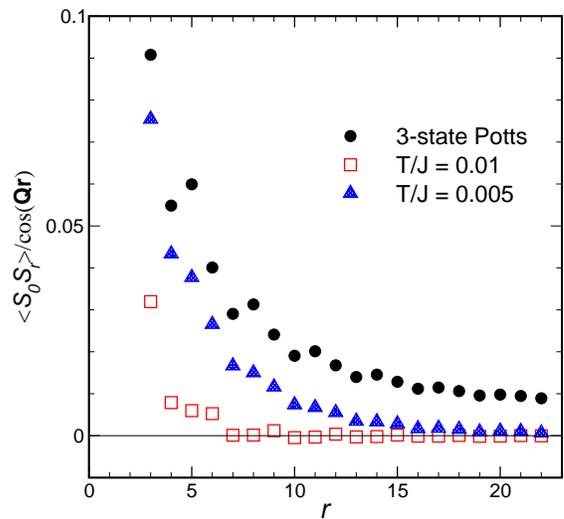}
\caption{(Color online) Distance dependence of 
spin correlators normalized to the correlations
of the $\sqrt{3}\times\!\sqrt{3}$ structure.
Correlators are measured along chain directions on the kagome i
lattice of linear size $L=48$. Distances are given in units of 
the lattice constant.
}
\label{correlationsK}
\end{figure}
% -------------------------------------------------------------------------

We have investigated spin correlations in the kagome antiferromagnet 
by using the following form of the staggered magnetization:
\begin{equation}
m_{\rm AF}^2 = \frac{6}{N^2} \sum_{l,i,j}\: 
\langle {\bf S}_{li}\cdot {\bf S}_{lj} \rangle \,
e^{i{\bf Q}({\bf R}_i-{\bf R}_j)} \ ,
\label{mAF}
\end{equation} 
where index $l$ numbers spins in the unit cell,
$i,j$ and ${\bf R}_{i,j}$ denote position of the unit cell
on the triangular Bravais lattice, and ${\bf Q} = (4\pi/3,0)$ 
is the wave-vector of the $\sqrt{3}\times\!\sqrt{3}$ structure.
The lattice constant, which is twice the intersite
spacing, is chosen as the length unit. The normalization factor
gives $m^2_{\rm AF}=1$ in the fully ordered structure.
The temperature dependence of the antiferromagnetic order parameter
for several lattice sizes is presented in Fig.~\ref{maf}.
Using the loop-flip algorithm \cite{huse92} we have also measured 
the antiferromagnetic order parameter for the three-state Potts 
antiferromagnet at $T=0$. The corresponding values are shown 
in Fig.~\ref{maf} by the horizontal arrows.

% -------------------------------------------------------------------------
\begin{figure*}[t]
\includegraphics[ width = 0.8\columnwidth ]{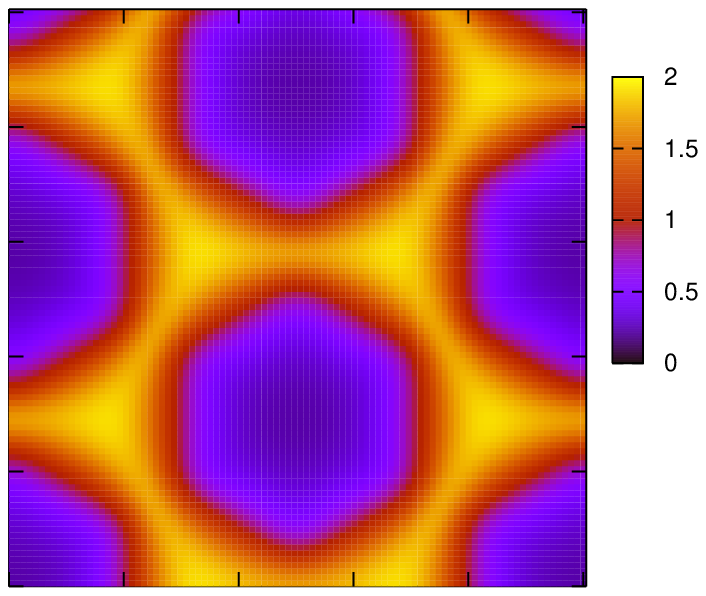}
\hspace{4mm}
\includegraphics[ width = 0.8\columnwidth]{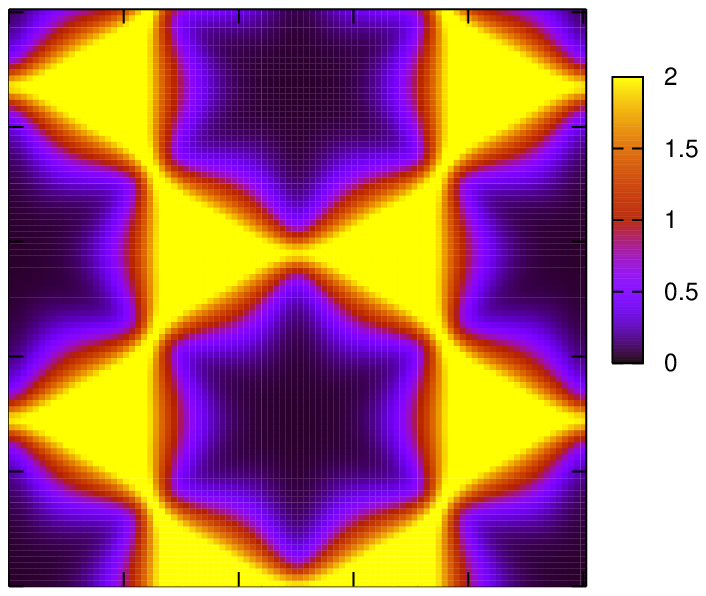}

\includegraphics[ width = 0.8\columnwidth]{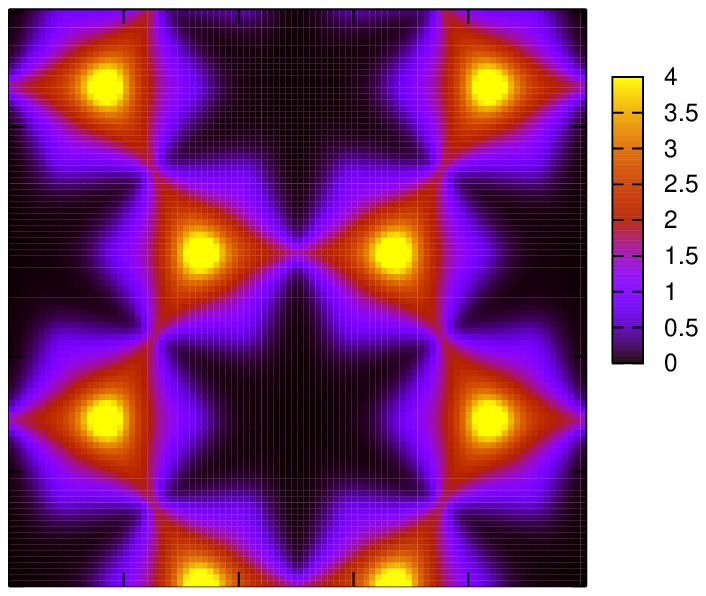}
\hspace{4mm}
\includegraphics[ width = 0.8\columnwidth]{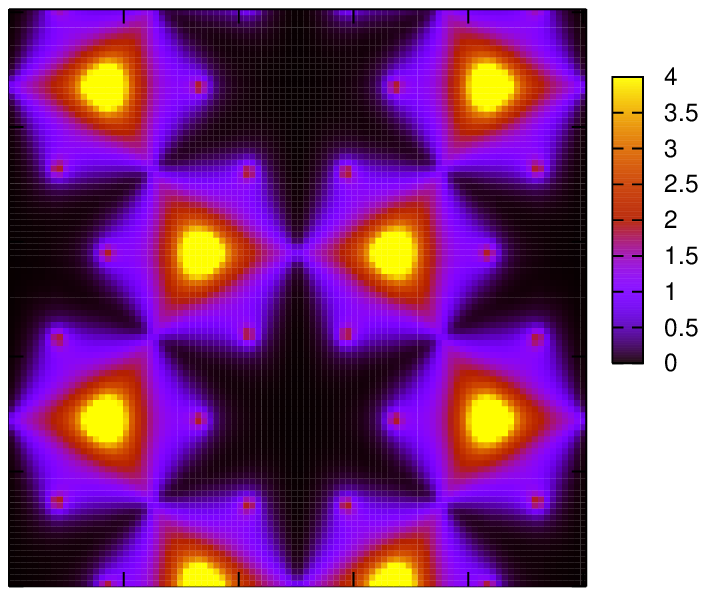}
\caption{(Color online) Magnetic structure factor of the kagome 
antiferromagnet in different temperature regimes. Region in the momentum
space corresponds to $0\leq q_{x,y}\leq 8\pi$.
Temperatures from left to right and 
from top to bottom are $T/J = 0.5$, 0.1, 0.02, and 0.005.
The intensity scales differ by a factor of 2 between the rows.
}
\label{maps}
\end{figure*}
% -------------------------------------------------------------------------

Spin correlations may also yield sharp signatures in the 
static magnetic structure factor $S({\bf q})$:
\begin{equation}
S({\bf q}) = \frac{1}{N} \sum_{i,j}\: 
\langle {\bf S}_{i}\cdot {\bf S}_{j} \rangle \,
e^{i{\bf q}({\bf r}_i-{\bf r}_j)} \ ,
\label{Sq}
\end{equation}
where ${\bf r}_i$ is a spin position in the 2D plane. 
The energy integrated neutron scattering cross-section
is proportional to the instantaneous spin correlator (\ref{Sq})
and provides an experimental tool for its measurement.
Previous numerical works on the kagome antiferromagnet
have considered only the powder averaged structure factor. \cite{reimers93,fak08}

In the whole range of temperatures the antiferromagnetic order 
parameter has much smaller values than the octupole or quadrupole moments,
see also Fig.~\ref{comparisonTQ}. 
The enhancement of $m^2_{\rm AF}$ observed
at $T/J \alt 0.01$ on small systems \cite{huse92,reimers93}
is significantly suppressed for larger clusters.
For every lattice size $L$ there is a characteristic
temperature below which the amplitude of the $\sqrt{3}\times\!\sqrt{3}$
correlations becomes larger than the corresponding 
correlations in the ground state of the three-state Potts antiferromagnet.
This happens because the $\pi$-folds dressed with short wave-length 
fluctuations acquire a finite linear tension,
whereas loops of all lengths are flipped with equal 
probability in the three-state Potts model.
Still, the enhancement relative to the three-state Potts model 
is progressively shifted to lower temperatures with increasing 
cluster size and in the end might be a finite-size 
effect.

Explicit comparison of spin correlations in the Heisenberg 
antiferromagnet and the three-state Potts model is presented 
in Fig.~\ref{correlationsK} for a lattice with $L=48$. 
A similar plot is given in Fig.~5 of Ref.~\onlinecite{huse92}. 
Once the difference in the length scales is taken into account,
the previously studied  cluster corresponds to $L=12$ in our notations.
At $T/J=0.005$, the spin correlations on the $L=12$ lattice
approach or even exceed the amplitudes of the Potts model, \cite{huse92}
which is consistent with the enhancement of the staggered
magnetization shown in Fig.~\ref{maf}. 
On the other hand, the correlations for the $L=48$ cluster
at the same $T$ fall consistently below 
the Potts model amplitudes, see Fig.~\ref{correlationsK}.
There is, therefore, a fundamental difference between 
the low-temperature behaviors of the spin tensor order parameters 
and $m_{\rm AF}$. While the octupole susceptibility
exhibits an exponential growth as $T\rightarrow 0$,
which signifies a finite value of the order parameter at $T=0$,
the staggered susceptibility $\chi_{\rm AF}$ has 
a much weaker increase, which for available system sizes
does not exceed corresponding values of the three-state Potts 
antiferromagnet.

Over the last several years it has been established that certain 2D and 3D classical
spin models governed by local constraints exhibit critical behavior
with power-law decay of spin correlations at long distances.
\cite{huse92,kondev95,garanin99,moessner03,huse03,isakov04,henley05}
A widely discussed consequence in application to the pyrochlore antiferromagnet
is a ``bow-tie'' shape of the magnetic structure factor 
with pinch
points in certain high-symmetry directions. \cite{zinkin97,isakov04,henley05}
Expansion around the large-$N$ limit for a 
$\mathrm{O(N)}$-symmetric spin model on the kagome lattice also yields 
a $1/r^2$ decay law for spin correlations with
similar bow-tie features in $S({\bf q})$. \cite{garanin99}
It is, therefore, interesting to compare
magnetic intensities in the three different temperature regimes
of the Heisenberg kagome antiferromagnet (Fig.~\ref{specific1})
with the above predictions.

The magnetic structure factor has been calculated for the kagome
lattice cluster with $L=24$ at four different temperatures.
The obtained results are shown in Fig.~\ref{maps} as intensity plots 
on a square slice in the reciprocal space with the extension
from $0$ to $8\pi$ in both directions. 
The highest temperature $T/J=0.5$ 
corresponds to the paramagnetic state, where thermal fluctuations
are strong and the magnetic intensity has only a broad structure in the 
momentum space.
In the cooperative paramagnetic state at $T/J=0.1$
the pinch points develop between triangular shaped
regions of strong intensity. In the vicinity of the pinch point $S({\bf q})$ 
has a nonanalytic form due to long-distance spin-spin correlations 
with dipolar-like angular anisotropy. \cite{isakov04,henley05,garanin99}
Upon cooling to a lower temperature $T/J=0.02$, the intensity
is redistributed in favor of centers of the triangular regions, 
which correspond to ${\bf Q}' = 2{\bf Q} = (8\pi/3,0)$ and equivalent
wave-vectors. These are not true Bragg peaks: 
as the spin correlations  fall off as $r^{-2}$ in this 
regime, \cite{garanin99} the peak intensity grows logarithmically
with the system size. 
Note, that peaks at the antiferromagnetic wave-vector $\bf Q$ and
equivalent positions are absent. 

Finally, when the coplanar correlations start to develop at $T/J=0.005$,
the narrow necks loose significantly in intensity, while new satellite peaks 
at ${\bf q} ={\bf Q}$ becomes noticeable. The intensity maps for $T/J = 0.1$, 
and 0.02 most closely 
resemble the analytic result for $S({\bf q})$ from the large-$N$ expansion. 
\cite{garanin99} However, the magnetic intensity exhibits
more structure once the order by disorder effect selects coplanar states.
The development of extra diffuse peaks 
in the low-temperature regime
is also observed in the powder averaged structure factor.
\cite{reimers93,schweika07,fak08}
Note, that the algebraic decay of spin correlations discussed 
above occurs at distances smaller than the correlation
length.
The inverse correlation length provides 
a natural width for all the nonanalytic features in the magnetic
structure factor $S({\bf q})$.

\section{Hyperkagome antiferromagnet}

\begin{figure}[t]
\includegraphics[ width = 0.6\columnwidth]{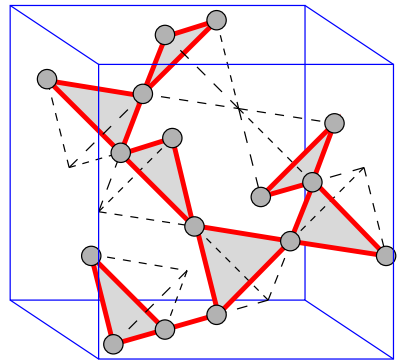}
\caption{(Color online) The hyperkagome lattice as a $1/4$-depleted 
pyrochlore structure.
}
\label{hyperkagome}
\end{figure}

The best known example of 3D geometrically frustrated lattice is
a network of corner-sharing tetrahedra of
the pyrochlore lattice, see, {\it e.g.}, Ref.~\onlinecite{moessner98}.
The experimental realizations 
include numerous magnetic pyrochlore and spinel 
compounds. The only example of a 3D lattice  
of corner-sharing triangles 
was so far provided by gadolinium gallium garnet 
Gd$_3$Ga$_5$O$_{12}$. \cite{kinney79,schiffer94} 
The recent experiment has found another interesting
example of a 3D triangular network of magnetic ions in 
Na$_4$Ir$_3$O$_8$. \cite{okamoto07} 
The corresponding lattice structure can be obtained
by $1/4$ depletion of the pyrochlore lattice such that
only three out of the four vertices of each tetrahedron are
occupied by magnetic ions, see
Fig.~\ref{hyperkagome}. By analogy with the kagome lattice,
this structure is called the hyperkagome lattice.
It contains 12 spins in the standard cubic unit cell.
The garnet lattice has, in contrast, 24 spins in its cubic unit cell
and consists of  two interpenetrating hyperkagome sublattices.
The positions of the atoms in each garnet sublattice are different from 
the 1/4-depleted
pyrochlore structure but topology of the two networks remains the same.
The local symmetry on magnetic sites is somewhat higher in the
garnet structure containing a two-fold rotational axis joining
centers of adjacent triangles.

% -------------------------------------------------------------------------
% -------------------------------------------------------------------------
In this Section we consider a classical Heisenberg 
antiferromagnet on the hyperkagome lattice.
The limited applicability of such a model to the above two 
magnetic materials has been mentioned before.
Still this model is quite interesting on its own, in particular,
to contrast it with the more familiar 2D system.
The arguments given before for the infinite degeneracy of
the classical ground states of the kagome antiferromagnet
(Sec.~I and II) are fully applicable to the 3D model
and will not be repeated here. The coplanar configurations
play again an important role at low temperatures.
The harmonic analysis finds $N/3$ zero-energy modes
for the hyperkagome antiferromagnet
out from the total $2N$ modes for $N$ classical spins.
\cite{mzh03,hopkinson07} This yields the same limiting
value for the specific heat $C/N|_{T\rightarrow 0}=11/12$ as for the 2D model.
Soft modes reside on closed loops, which pass through at least ten triangles.
\cite{mzh03,lawler08}

% -------------------------------------------------------------------------
\begin{figure}[t]
\includegraphics[ width = 0.95\columnwidth]{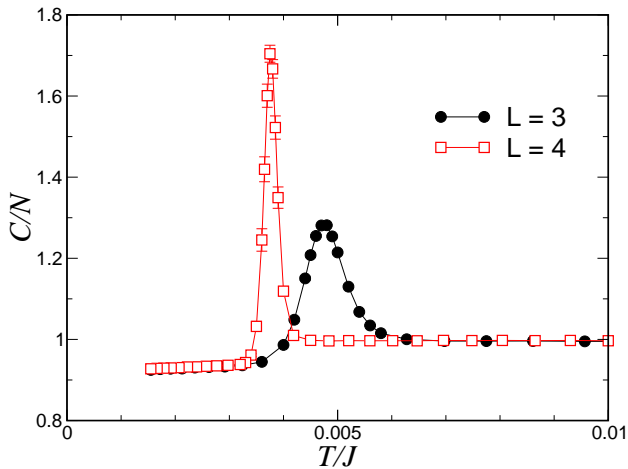}
\caption{(Color online) Low-temperature specific heat
of the classical hyperkagome antiferromagnet.
}
\label{specific3}
\end{figure}
% -------------------------------------------------------------------------

A submanifold of coplanar configurations is again mapped
to the ground states of the three-state Potts antiferromagnet
on the hyperkagome lattice.
The precise number of such states or, equivalently, 
the ways of coloring sites with three colors is not known
for the 3D lattice.
A local gauge representation of the classical constraint
with the Maxwellian-type action predicts 
the dipolar form of the spin-spin correlation function. 
\cite{huse03,hopkinson07}  Performing loop-algorithm
simulations in the ground state ensemble of the Potts antiferromagnet
we indeed found a very fast decay of spin correlations consistent
with an $r^{-3}$ dependence.

The general mode counting arguments suggest the presence
of the order by disorder
effect  for three-component spins on a lattice of
corner-sharing triangles. \cite{moessner98} 
Though the early numerical work found no
evidence of such an effect, \cite{petrenko00}
the  recent Monte Carlo study detected a 
first-order transition at $T/J\alt 0.004$. \cite{hopkinson07}
The authors argued in favor
of a nematic state below the transition.
As we have shown in Sec.~II the full symmetry breaking
pattern in the submanifold of the coplanar states
is described by the octupolar order parameter.
The first-order nature of the transition
is not surprising in this respect as the renormalization group
analysis of the Landau free-energy functional (\ref{FT})
finds no stable fixed-point solution below four dimensions. \cite{radzihovsky01}

To verify the formation of the octupolar ordering
in the hyperkagome antiferromagnet
we have performed the classical Monte Carlo simulations on
periodic clusters with $N=12L^3$ spins and $L=3$--$6$.
The equilibration problem at low temperatures 
in three dimensions becomes much more severe compared to the 2D model.
The published data \cite{hopkinson07} show, for example, 
a strong hysteresis for $L\geq 4$, which clearly indicates
that large clusters fall out of equilibrium.
We resort, therefore, to the exchange Monte Carlo
algorithm \cite{hukushima96} in a conjunction with the hybrid updates
described before. 
We use between 30 to 50 replicas depending on cluster size
distributed in the temperature interval
$0.001<T/J<0.02$. All replicas are initiated with random spin configurations,
which are equilibrated for $10^6$ exchange MC steps.
These are followed by measurements for $5\times 10^5$ MC steps.
Finally, the results have been averaged over 20 independent runs
to determine statistical errors. 

% -------------------------------------------------------------------------
\begin{figure}[t]
\includegraphics[ width = 0.95\columnwidth]{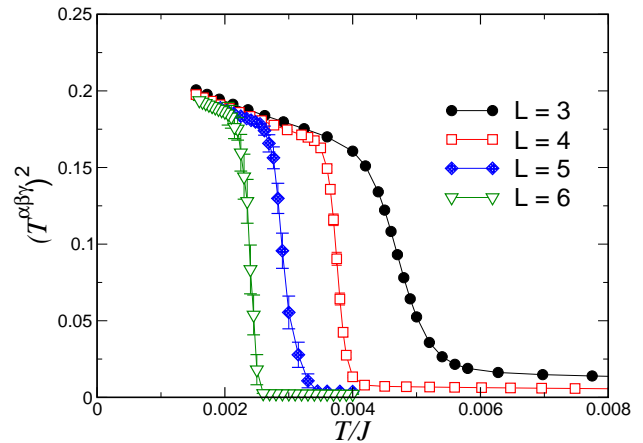}
\caption{(Color online) Mean-square of the octupolar
order parameter versus temperature for the hyperkagome antiferromagnet.
}
\label{opGGG}
\end{figure}
% -------------------------------------------------------------------------

The numerical data for the specific heat obtained on two smallest clusters 
are presented in Fig.~\ref{specific3}. At temperatures 
below the anomaly, the specific heat approaches the value 
$C/N=11/12$ predicted for the coplanar state by the mode counting analysis.
The fast growth of the peak height is consistent with $C/N \propto L^3$
scaling expected for the first-order transition. \cite{binder84}

The temperature dependence of the mean square of the octupolar order
parameter is shown in Fig.~\ref{opGGG}.
The transition temperature for each cluster can be estimated at
the mid-height of the jump.
The order parameter exhibits extremely weak finite size dependence at
low temperatures, which signifies development of the true long-range
octupolar ordering in this 3D spin model.
Still, a certain amount of disorder in the low-temperature
phase is evidenced by deviations of 
$\langle (T^{\alpha\beta\gamma})^2\rangle$  from the limiting value of 
1/4 for the fully ordered octupolar structure.
A simple scaling of the transition temperature with the cluster size
yields $T_c/J = 0.002\pm0.0003$ for the transition temperature.
In order to locate more accurately the  point of first-order transition,
one needs to simulate bigger lattices, which 
appears to be a difficult task even for the exchange Monte Carlo
algorithm.

% -------------------------------------------------------------------------
\begin{figure}[t]
\includegraphics[ width = 0.9\columnwidth]{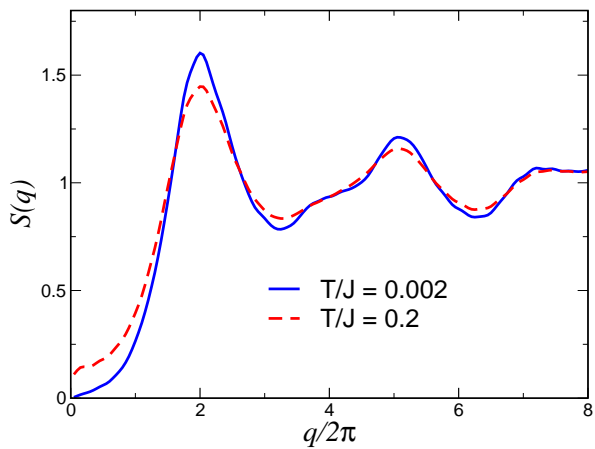}
\caption{(Color online) Powder averaged magnetic structure factor
for the hyper\-kagome antiferromagnet.
Wave-vectors are given in reciprocal lattice units.
}
\label{sqGGG}
\end{figure}
% -------------------------------------------------------------------------

In order to check a tendency to a long-range
magnetic ordering below $T_c$ at {\it a priori}
unknown wave vector we have calculated the angular-averaged
structure factor:
\begin{equation}
S(q) = \frac{1}{N} \sum_j \langle {\bf S}_i\cdot{\bf S}_j\rangle\;
\frac{\sin(q|{\bf r}_{ij}|)}{q|{\bf r}_{ij}|} \ .
\end{equation}
The neutron diffraction experiments on polycrystalline samples 
allow to directly measure $S(q)$.
The results for the $L=5$ cluster are presented 
in Fig.~\ref{sqGGG}.
The static structure factor $S(q)$ below $T_c$ exhibits broad diffuse peaks 
with no sharp features, which could point at a long-range order.
We have found only a tiny change in $S(q)$ across the first-order 
transition. Variation of temperature by 2 orders of magnitude from 
$T/J=0.2$ to $0.002$ leads only to a small increase $\sim 15$\% in the main peak
intensity. A similar form of the diffuse structure factor
at high temperatures was previously obtained 
in the numerical work on gadolinium gallium garnet.\cite{petrenko98}
The nearest-neighbor classical Heisenberg antiferromagnet on the hyperkagome 
lattice provides a unique example of the long-range ordered state
of magnetic octupoles with algebraically decaying spin-spin correlations.

Returning back to Na$_4$Ir$_3$O$_8$, the recent experiment\cite{okamoto07}
has demonstrated the absence of magnetic ordering down to 2~K,
which is significantly smaller
than the scale of antiferromagnetic interactions deduced from
the Curie-Weiss constant $\theta_{CW} \sim 650$~K.
Such a behavior is consistent with strong geometrical frustration
found for the nearest-neighbor hyperkagome antiferromagnet. 
\cite{petrenko00,hopkinson07,lawler08}
It would be interesting to compare 
the neutron diffraction data in the spin-liquid state of  
Na$_4$Ir$_3$O$_8$ 
with the above results for $S(q)$. This should allow one, in particular, 
to verify relevance of the nearest-neighbor Heisenberg model
to the real material.

\section{Conclusions}

We have clarified in the present work that selection
of the coplanar states in the kagome antiferromagnet 
is properly described by the development of the octupolar
(third-rank spin tensor)
order parameter. Our Monte Carlo simulations yield
$T_k \approx 0.004J$ for the onset of coplanar ordering, which
is lower than the previous estimates. \cite{chalker92,reimers93}
Furthermore, we suggest that $T_k$ may correspond to a topological
transition, which consists in unbinding of fractional vortices.
Presence of these topologically stable point defects follows
from the nontrivial degeneracy space of the octupolar order parameter.
The antiferromagnetic $\sqrt{3}\times\!\sqrt{3}$ correlations 
are also enhanced at low temperatures, though the corresponding  correlation
length remains significantly shorter than 
the characteristic length scale for octupolar correlations. 
Our MC data for big lattices with $L\geq 36$
do not confirm the previously made suggestion \cite{huse92}
that the asymptotic $\sqrt{3}\times\!\sqrt{3}$ ordering develops 
in the limit  $T\rightarrow 0$. Precise numerical study
of the lowest temperature region $T/J\alt 10^{-3}$ definitely requires
simulations on clusters with $L>72$, which is impossible
without new Monte Carlo algorithms (but see also the last paragraph).

The classical hyperkagome antiferromagnet provides a unique
example of the spin model with the long-range octupolar ordering.
In the broken-symmetry state 
below a fluctuation induced first-order transition
spin correlations
remain critical with a power-law decay $r^{-3}$ at long distances.
\cite{huse03,hopkinson07}
It is interesting to study the phase diagram in magnetic field
for this model, in particular, in relation to the experimental
diagram of 
Gd$_3$Ga$_5$O$_{12}$\, 
which exhibits a field-induced ordered phase.
\cite{kinney79,schiffer94,petrenko98} 

Let us finish with a few comments on the possible relation of 
the topological transition to the spin-glass like behavior observed
in some jarosites. \cite{wills98,wills00} 
Similar to the previous numerical studies we find a very strong
tendency to spin freezing once the coplanar configurations
are stabilized at $T<T_v$.
It is the hybrid Monte Carlo algorithm adopted in the present
work, which allows one to equilibrate large spin systems 
at low temperatures. The spin reshuffling dynamics of the microcanonical 
sweeps has no simple analog in real magnetic materials. 
The kagome antiferromagnet will be, therefore, stuck in one of the many
degenerate coplanar states with frozen structure of chirally 
ordered domains.
Rotation of spins along spin folds is suppressed due to development
of free-energy barriers of the entropic origin. 
\cite{chalker92,ritchey93,chandra93}
Above the topological transition, when thermally excited vortices
destroy the common plane and remove the entropic barriers,
the single spin-flip dynamics becomes effective again.
The above ideas for the unconventional spin-glass transition
were pioneered by Ritchey and co-workers. \cite{ritchey93,chandra93}
The results of our work put emphasis on a hidden 
topological transition behind a spin-glass freezing in geometrically
frustrated magnets. 
Though the corresponding temperature scale comprises
only a small fraction of $\theta_{CW}$ for the Heisenberg antiferromagnet,
intrinsic $XY$ anisotropies in real materials can significantly
enhance $T_v$ by transforming it to the Kosterlitz-Thouless
transition for 1/3-vortices.

\section*{Acknowledgments}

I am grateful to B. F\aa k, R. Moessner, and P. Thalmeier 
for helpful discussions.
I thank the MPI-PKS (Dresden) for the hospitality.

\end{document}